\documentclass[%
 reprint,
 amsmath,amssymb,
 aps,
]{revtex4-1}

\usepackage{graphicx}% Include figure files
\usepackage{dcolumn}% Align table columns on decimal point
\usepackage{bm}% bold math
\usepackage{braket}

\begin{document}

\preprint{APS/123-QED}

\title{Temporally Asymmetric Bi-photon States in Cavity Enhanced Optical Parametric Processes}% Force line breaks with \\

\author{Usman A. Javid$^{1,2}$}
\author{Steven D. Rogers$^{2,3}$}%
\author{Austin Graf$^{1,2}$}%
\author{Qiang Lin$^{1,2,4}$}%
\email{qiang.lin@rochester.edu}
\affiliation{$^{1}$Institute Of Optics, University of Rochester, Rochester NY 14627, USA}
\affiliation{$^{2}$Center for Coherence and Quantum Optics, University of Rochester, Rochester NY 14627, USA}
\affiliation{$^{3}$Department of Physics and Astronomy, University of Rochester, Rochester NY 14627, USA}
\affiliation{$^{4}$Department of Electrical and Computer Engineering, University of Rochester, Rochester NY 14627, USA}

\date{\today}% It is always \today, today,

\begin{abstract}
Generation and control of quantum states of light on an integrated platform has become an essential tool for scalable quantum technologies. Chip scale sources such as nonlinear optical microcavities have been demonstrated to efficiently generate entangled bi-photon states. However these systems have little control over the continuous variable time-energy entanglement of the photons. We demonstrate such control by preparing bi-photon states with asymmetric temporal wavefunctions by selectively modifying the density of states of the cavity modes taking part in the interaction using Rayleigh scattering-induced strong coupling of optical modes of a resonator. These states reveal exotic coherence properties and show a path forward for continuous variable quantum state engineering on a chip.
\end{abstract}

\maketitle

%\tableofcontents

\section{Introduction}
Behaviors of natural systems with quantum mechanical degrees of freedom have proven to be useful not just for better understanding the physical world but for their impact in areas such as computing \cite{comp}, communication \cite{comm}, simulation \cite{sim}, security \cite{cryp1,cryp2}, and metrology \cite{met1,met2}. Such applications invariably rely on generation of quantum states and their control. Efforts have been undertaken to precisely engineer these systems to create sources of novel quantum states. Among these systems, photons have emerged to be of great promise as a carrier of quantum information \cite{phot1,phot2,phot3,phot4,phot5,phot6} owing to their robustness against noise, ease of control and versatile set of degrees of freedom to encode quantum information such as polarization, path, spatial mode, time and frequency.

\begin{figure*}
  \includegraphics[scale=0.97]{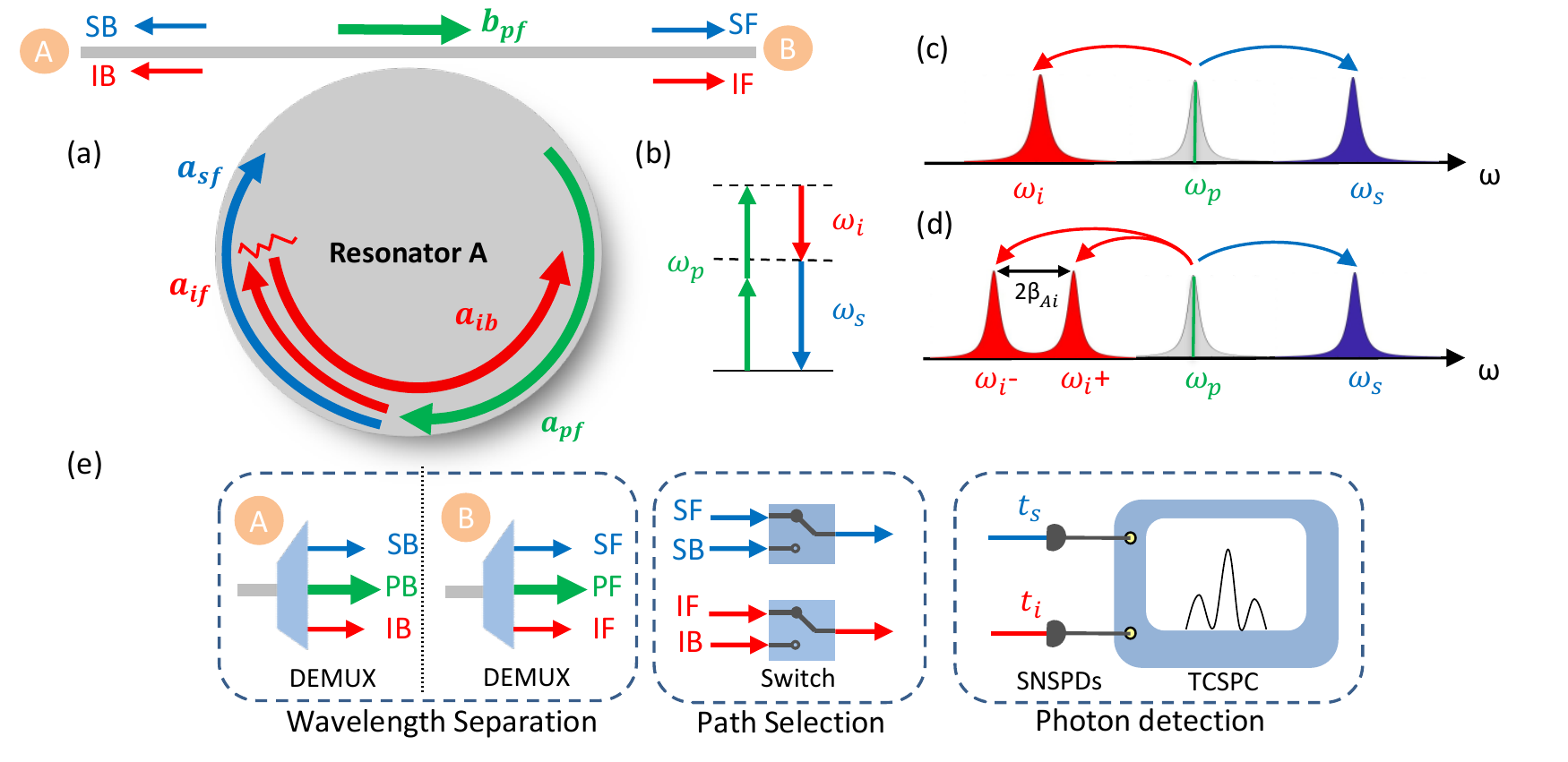}
  \centering
  \caption{Illustration of the device physics and photon pair creation pathways in a silicon WGM microdisk resonator labeled as resonator A. (a) A pump laser propagating in a waveguide in the forward direction labelled as $b_{pf}$ is coupled into the resonator by bringing it into the resonator's near field. The circulating pump inside the cavity generates photon pairs into the signal mode (blue) and idler mode (red) with spontaneous four wave mixing (SFWM). All propagating fields inside the resonator and their corresponding resonance frequencies are labeled as $a_{jk}$  and $\omega_j$ where j = p (pump), s (signal), i (idler) and k = f (forward/clockwise), b (backward/counter-clockwise). The idler mode has Rayleigh scattering-induced coupling of the forward and backward propagating modes with a coupling rate $\beta_{Ai}$. The photons are coupled out with the same waveguide both in the forward and backward direction labelled as JK. (b) Energy level diagram for SFWM. (c) Spectral profile of a resonator without any coupling mechanism. All the three modes participating in SFWM are engineered to be equally spaced in frequency and have a lorentzian lineshape. (d) Spectral profile for resonator A. Due to the coupling, the idler mode forms dressed states and the resonance takes a 'doublet' lineshape while the pump and signal remain the same. (e) Experimental setup for the characterization of the bi-photon state. The photons are separated with telecom band demultiplexers and then the path selection is done with optical switches. The photons can be detected in four configurations SFIF, SFIB, SBIF, SBIB. DEMUX: Demultiplexer. SNSPDs: Superconducting nanowire single-photon detectors. TCSPC: Time-correlated single-photon counter. $t_s$ and $t_i$: Signal and idler arrival times.}
  \label{device}
\end{figure*}

\par
One of the most efficient methods of generating quantum states of light is based on optical parametric processes such as spontaneous parametric down conversion (SPDC) and spontaneous four wave mixing (SFWM). These sources generate time-energy entangled photon pairs and have been demonstrated to also produce entanglement in discrete variables such as polarization and path \cite{pol1,pol2,rabi}, spatial mode \cite{space1,space2} , time and frequency bins \cite{TF1,TF2} etc. Efforts have been made to engineer the continuous variable time-frequency entanglement that emerges from these processes, for example, by manipulating pump properties, engineering dispersion of the nonlinear medium to obtain exotic phase matching conditions or by time or frequency domain post-processing of the generated photons \cite{TFeng1,TFeng2,TFeng3,TFeng4,TFeng5,TFeng6,TFeng7,TFeng8,TFeng9}. Cavity confinement of these parametric processes greatly enhances the efficiency and purity of the generated bi-photon states as well as increases their spectral brightness by reducing the spectrum of the photons to the lineshape of the cavity resonance. This, however, severely limits any manipulation that can be done to the time-frequency entanglement of the state.

\par
Here we report an efficient approach to flexibly control the spectro-temporal properties of a bi-photon state in a high-quality microresonator, via independent engineering of the densities of the states of the interacting modes. We demonstrate generation of bi-photon states with asymmetric temporal wavefunctions generated by selectively modifying only one of the two signal and idler densities of states. These wavefunctions reveal starkly different behavior in the second order cross-correlation function $g^{(2)}_{si}(\tau)$ ($s$ = signal, $i$ = idler) before and after the triggering off one of the photons with sharp cusps forming at the triggering instant.

We realize this process using an on-chip ultra-high-Q silicon microdisk resonator. The third-order nonlinear susceptibility of silicon allows SFWM in three adjacent cavity resonances, which are dispersion engineered to be equally spaced in frequency. Due to the rotational symmetry of the device, each cavity resonance exhibits two degenerate cavity modes in which the light propagates clockwise and counter-clockwise, respectively. We utilize Rayleigh back-scattering \cite{scatter} to introduce coupling between these two counter-propagating modes, which lifts their degeneracy and re-normalizes them into dressed states. Selectively engineering the coupling strength of Rayleigh scattering enables individual control of the density of states of signal and idler photons by controlling the frequency splitting of the dressed modes. To show this capability, we investigate two resonators labelled as A and B. Resonator A has such coupling in one out of the three cavity modes taking part in the interaction creating an asymmetry in the spectral domain. This leads to an asymmetric temporal profile of the bi-photon state. Resonator B has identical mode splitting in both the signal and idler modes, thus maintaining the symmetry. These two devices exhibit remarkably distinctive coherence properties of the emitted light, and we observe oscillatory revivals in the second order quantum coherence function of the individual photon channels $g^{(2)}_{jj}(\tau)$ (j = s, i) when the symmetry is maintained.

\begin{figure*}
\centering
  \includegraphics[scale=0.99]{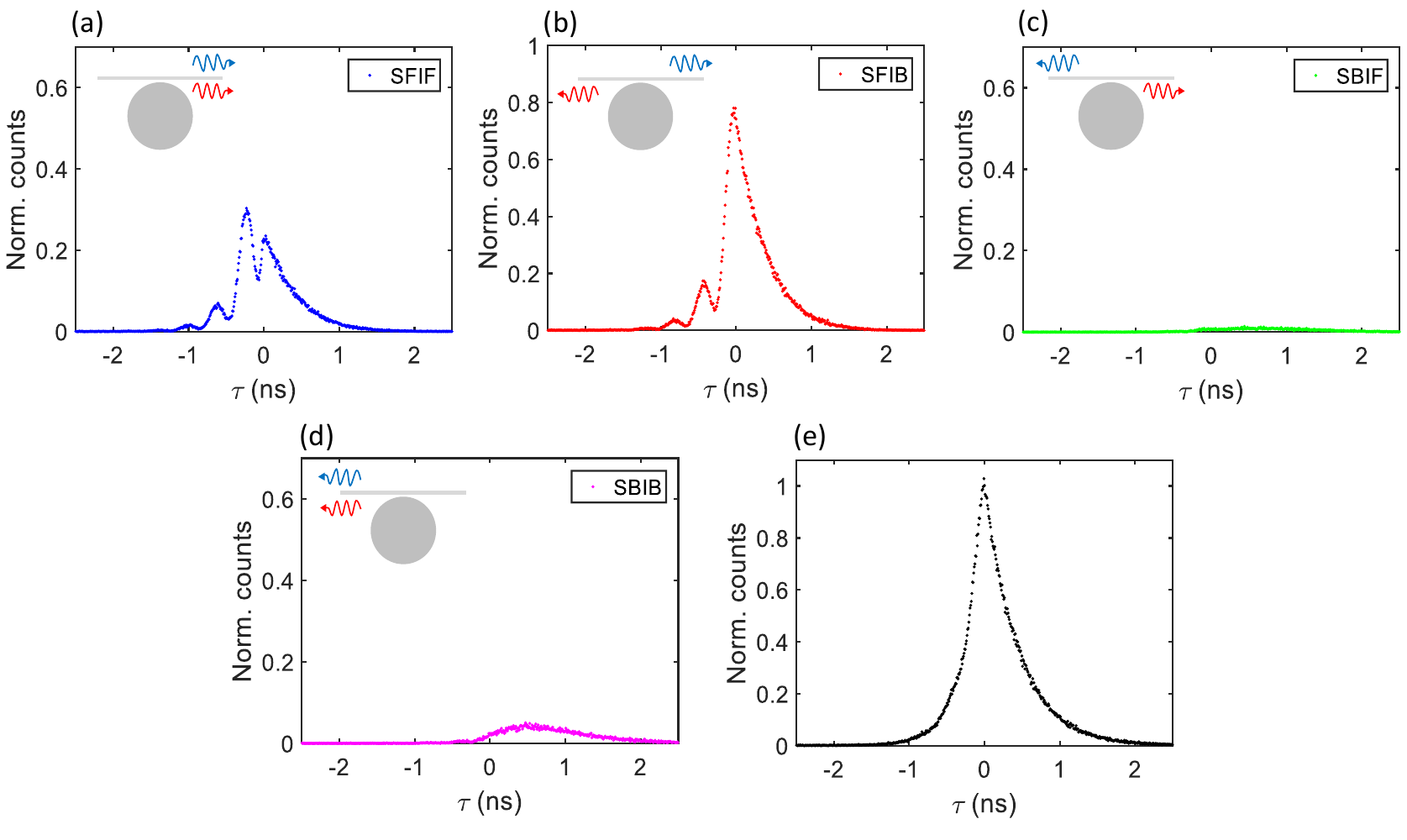}
  \caption{Generation of asymmetric temporal waveforms of the bi-photons verified by measuring the second-order cross-correlation function $g^{(2)}_{si}(\tau)$ ($\tau=t_s-t_i$) plotted for all four detection configurations. The plos are normalized to the peak of the coincidence envelope in (e). (a) Signal forward - idler forward (SFIF). (b) Signal forward - idler backward (SFIB). (c) Signal backward - idler forward (SBIF). (d) Signal backward - idler backward (SBIB). (e) The four detection configurations add to give an exponential decaying envelop of the cross-correlation.}
  \label{fig2}
\end{figure*}

\section{Results}
\subsection{Device Physics}
The resonators in question are etched from a silicon on insulator wafer with a radius of 4.5 um and a thickness of 260 nm. Figure \ref{device}(a) illustrates the device physics for resonator A. It has three adjacent modes, equally spaced in frequency. Light is coupled into the central mode in the forward direction (labeled as $\hat{b}_{pf}$). The propagating modes inside the cavity are labelled as $\hat{a}_{jk}$  where j = p (pump), s (signal), i (idler) and k = f (forward/clockwise), b (backward/counter-clockwise). Ordinarily, in the absence of any coupling mechanism, photons are generated in the same direction as the travelling pump mode with a combination of frequencies that conserve energy (Fig. \ref{device}(b)) and there is only one photon creation pathway as show in Fig, \ref{device}(c). In the presence of Rayleigh scattering, the light inside the resonator splits into the two counter-propagating modes $\hat{a}_{jf}$ and $\hat{a}_{jb}$. Figure \ref{device}(d) shows the asymmetric cavity mode structure of resonator A and the photon creation pathways. The quasi-TM modes of this device have average intrinsic quality factors of 1.3 million with center wavelengths 1527.1 nm, 1545.4 nm and 1564.1 nm for the signal, pump and idler mode respectively (see Appendix A for device characterization details). The idler mode has a splitting $\beta_{Ai}/\pi$ = 2.52 GHz and a loaded linewidth $\Gamma_{Ai}/2\pi$ = 595 MHz putting it in the strong coupling regime with a linewidth to splitting ratio $\beta_{Ai}/\Gamma_{Ai}$ = 2.12. The signal mode has a small coupling creating a splitting $\beta_{As}/\pi$ = 150 MHz with a loaded linewidth $\Gamma_{As}/2\pi$ = 269 MHz which puts it in the weak coupling regime with $\beta_{As}/\Gamma_{As}$ = 0.28. The signal splitting is not visible when the device is decently coupled to an external waveguide (a tapered optical fiber in this case) and thus behaves like a singlet resonance. Appendix A gives more details on the signal resonance characterization. The pump mode has no measurable splitting. Since the idler mode has forward-backward coupling, it will create a standing wave while the signal photon will dominantly remain a travelling wave mode in the same direction as it is created. Since the mode coupling of the idler exceeds the loaded cavity linewidth, energy conservation forces idler photons to be generated in only one of the two dressed states. This is further investigated in section \ref{sec2}. Figure \ref{device}(e) depicts the experimental setup to characterize the bi-photon state. The photons are coupled out into the tapered fiber and separated with telecom band multiplexers. Then the path is selected using optical switches. The photons are captured by a pair of superconducting nanowire single photon detectors (SNSPD) connected to a coincidence counter. The detection system’'s total timing jitter is approximately 45 ps; far shorter than the temporal features of interest. Therefore the bi-photon coherence can be resolved.

\begin{figure*}
  \includegraphics[scale=0.9]{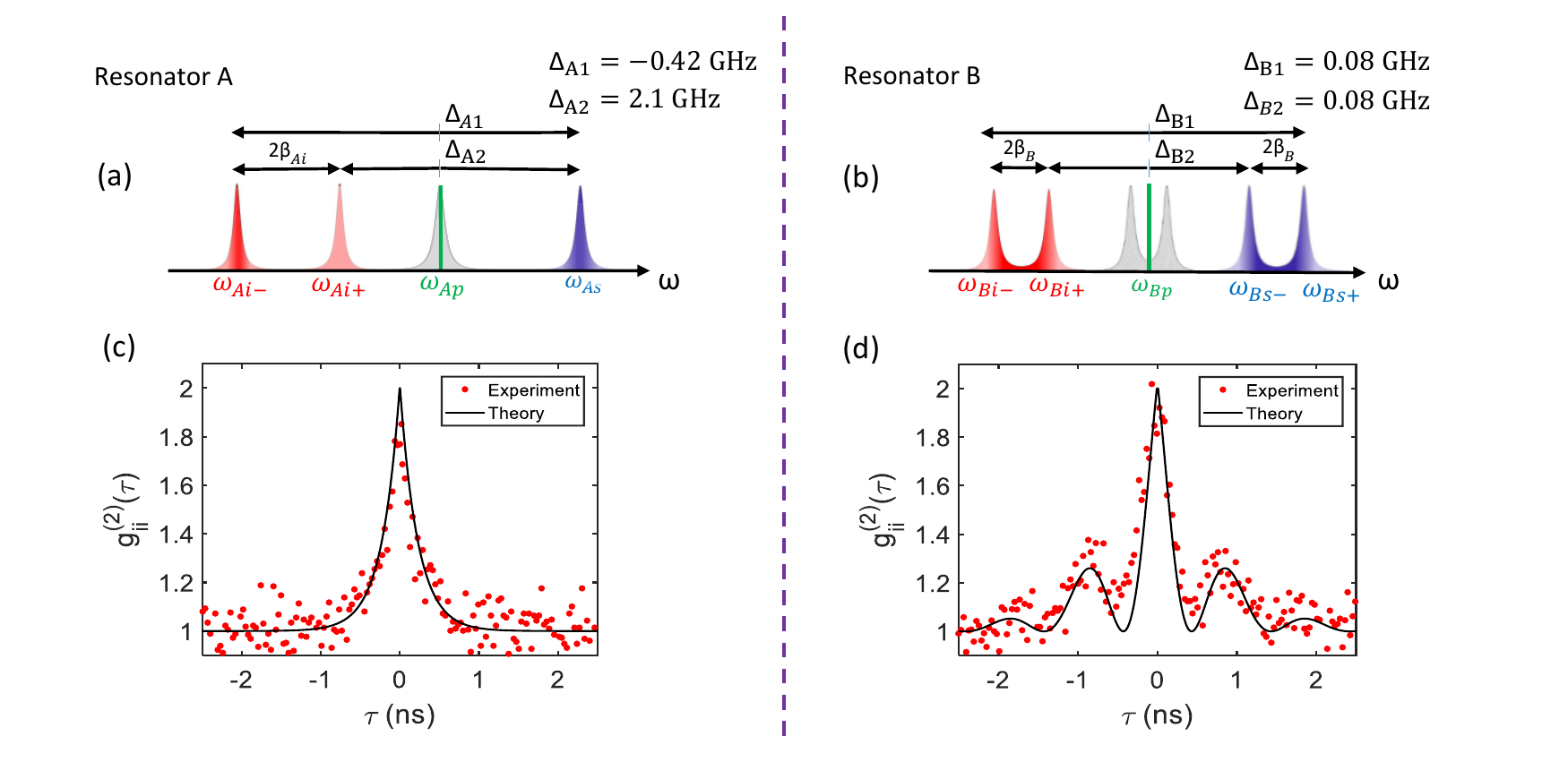}
  \centering
  \caption{Comparison of frequency matching and single channel coherence properties for Resonators A and B. Frequency separation between adjacent modes is measured to calculate energy mismatch parameters defined in Eqs. (\ref{eq1}) to (\ref{eq4}) for (a) resonator A and (b) resonator B. Second order self-correlation function $g^{(2)}_{ii}(\tau)$ for the forward propagating idler (IF) channel for (c) resonator A and (d) resonator B.}
  \label{fig3}
\end{figure*}

\subsection{Bi-photon Coherence Properties} \label{sec2}
Figure \ref{fig2} plots the second order cross-correlation $g^{(2)}_{si}(\tau)$ of the signal and idler fields where $\tau=t_s-t_i$ is the difference in arrival time of the two photons. The plots are labeled with their corresponding detection configuration IJ (I, J = F, B). The measurements are made with the laser tuned to the center of the pump mode. The plots show exponential decay of the correlation associated with a Lorentzian lineshape which abruptly switches to coherent oscillations at $\tau$ = 0. The exponential decay envelope corresponds to the linewidth of idler resonance for $\tau<0$ and signal for $\tau>0$ while the oscillation frequency matches with the mode splitting $2\beta_{Ai}$. The emergence of oscillations is due to the forward-backward scattering of the idler photon inside the resonator. For coincidences at $\tau<0$, the signal photon is detected first and the idler photon can either be detected in the forward or backward channel depending on the instant it leaves the resonator into the coupling fiber. This gives rise to the observed oscillations. On the other hand, for coincidences at $\tau>0$, the idler photon is detected first, and since the scattering rate of the signal is small, it has a high probability to exit the cavity in the forward direction and there are no oscillations in this part of the coherence function.

\par
Figure \ref{fig2}(b) plots the signal forward - idler backward configuration where the oscillatory features are complimentary to Fig. \ref{fig2}(a) creating time-evolving path-encoded states of the two photons \cite{rabi}. The coincidence flux in the remaining two configurations (Fig. \ref{fig2}(c), (d)) is considerably low compared to the configurations where the signal is detected in the forward direction indicating that most of the bi-photons are generated in the forward direction as expected. Ideally, the configurations where the signal is detected in the backward direction should not have any coincidences. The counts in these configurations occur because the signal is not a true singlet and does have a weak forward-backward coupling creating a small probability of its photons scattering backwards within the cavity lifetime. Figure \ref{fig2}(e) plots the sum of the coincidence traces of the four detection configurations giving the exponential decay envelope expected from a travelling wave resonator.

\par
The modification to the spectrum of the generated light leaves its signature on the coherence properties of individual photon channels as well. Spontaneous parametric processes such as SPDC and SFWM generate photons with thermal statistics \cite{Mandel-Wolf}, as the second order coherence of the generated fields shows bunching with $g^{(2)}_{jj}(\tau)$ (j = s, i) monotonically decaying to 1. It is known that light generated within coherently coupled optical modes can generate oscillatory revivals in the second order coherence (for example see \cite{blockade}). The strong coupling between counter-propagating optical modes in this system offers such a mechanism. The measurement is made with the forward propagating idler mode split with a 50:50 coupler and measured with the same detection setup as Fig. \ref{device}(e). The results are plotted in Fig. \ref{fig3}(c). The data shows thermal statistics as expected, however there are no new features. It must be noted that the coherence function is dependent on the spectral profile of the mode, which in turn depends on the laser detuning and the energy matching bandwidth. In order to investigate this further, we consider resonator B. This device has a loaded cavity linewidth of 174 MHz for the signal and 251 MHz for the idler mode and a coupling rate $\beta_{B}/\pi$ = 1 GHz for both the modes (see Appendix A for details). When pumped near the zero detuning point, the energy mismatch for each photon pair combination needs to be measured. We define two parameters for each device
\begin{align}
\label{eq1}
  \Delta_{A1}&=(\omega_{As}-\omega_{Ap})-(\omega_{Ap}-\omega_{Ai-})\\
  \label{eq2}
  \Delta_{A2}&=(\omega_{As}-\omega_{Ap})-(\omega_{Ap}-\omega_{Ai+})\\
  \label{eq3}
  \Delta_{B1}&=(\omega_{Bs+}-\omega_{Bp})-(\omega_{Bp}-\omega_{Bi-})\\
  \label{eq4}
  \Delta_{B2}&=(\omega_{Bs-}-\omega_{Bp})-(\omega_{Bp}-\omega_{Bi+})
 \end{align}
where the subscripts + and - indicate higher and lower energy dressed states respectively. This measurement is made by scanning a tunable laser through the cavity, identifying the mode locations while simultaneously passing the same light through a fixed unbalanced Mach-Zehnder interferometer with a known free spectral range. By counting the interference fringes, we have measured the parameters in Eqs. (\ref{eq1}) to (\ref{eq4}). The results are shown in Fig. \ref{fig3}(a) and (b).  It is clear that the idler channel of only resonator B has a doublet lineshape since resonator A has an energy mismatch of 2.1 GHz for the dressed mode $\omega_{Ai+}$, far exceeding the cavity linewidth. Therefore this mode will not be populated. In contrast, the energy mismatch parameters for resonator B are at 80 MHz, well within the linewidth of the resonances. We measure the second-order self-correlation of the forward propagating idler channel for resonator B. The results are shown in Fig. \ref{fig3}(d). Here we observe oscillatory revivals of coherence with a period just below 1 ns, matching with the mode splitting for resonator B, while the same measurement for resonator A revealed no such oscillations. We can theoretically investigate this behavior of the coherence function, without solving the complete system, by considering the spectrum of the idler photons of the two resonators. It is known that for a chaotic light source such as this, the second order coherence function can be obtained from its corresponding first order coherence function as

\begin{align}
    g^{(2)}(\tau)=1+|g^{(1)}(\tau)|^2\\
    g^{(1)}(\tau)=\mathcal{F}\mathcal{T}^{-1}[|S(\omega)|^2]
 \end{align}
 where $S(\omega)$ is the spectrum of the light and $\mathcal{F}\mathcal{T}^{-1}$ represents an inverse Fourier transform. Without any loss of generality, we can define the spectrum for a doublet resonance as
\begin{align}
    S(\omega)=\frac{1}{\gamma_t/2-i\omega} \otimes [\delta(\omega-\beta)+\delta(\omega+\beta)]
 \end{align}
 where the modulus square of $\frac{1}{\gamma_t/2-i\omega}$ is a lorentzian with a linewidth $\gamma_t$. $\beta$ is the mode coupling rate and $\otimes$ represents a convolution operation. The calculated results are plotted as black solid lines in Fig. \ref{fig3}(c) and (d). The actual spectral profile of the idler mode is dependent on the signal mode as well as the pump and the dispersion. However as long as the centers of both resonances and their linewidths are well matched, this model should give a good approximation. Indeed, we see good agreement between theoretical and experimental results. It is important to note that although the idler mode has a coherent forward-backward coupling in both the resonators, the oscillations in the case of resonator A are suppressed due to the effect of the signal mode on the idler since energy conservation limits the spectrum of the idler to only one of the two dressed modes and it effectively behaves like a singlet resonance. We also note that the difference of the pump mode between the two resonators is not the reason for this behavior. The pump mode being a doublet allows the photons to be generated in both directions in resonator B changing only the initial state of the photons compared to resonator A, where the photons are generated only in the forward direction. Since both resonators' idler modes have a coherent forward-backward coupling, both will have counter-propagating idler photons inside them.

\section{Conclusion}
This work demonstrates generation of bi-photon states with asymmetric entanglement profiles using cavity enhanced SFWM by selectively modifying the density of states of the cavity modes involved in the four-wave interaction. This behavior simultaneously manifests signatures of strong and weak coupling occurring through Rayleigh back-scattering between counter propagating cavity modes. Although, we rely on nano-scale roughness of the resonators (present due to fabrication imperfections) to induce this coupling, the coupling strength can be precisely controlled and made highly selective of the resonator modes \cite{SMS}. Together with control over external coupling rates and pump detuning, this gives control over the spectral and temporal profile of the nonlinear interaction and the resulting bi-photon state. This system can also be used to engineer single photon wavepackets by heralding one photon conditioned on a trigger off of the other. The heralded photon will fall into a pure state if the triggering resolution is much finer than the bi-photon coherence time \cite{Du}, a condition met by our detection system. We envision that such linear modifications to nonlinear interactions will lead to generation of quantum states of light with exotic coherence properties and benefit applications in continuous variable time-frequency domain quantum information and computing by engineering \cite{TF-QC} entanglement of bi-photon and multi-partite states.

\begin{acknowledgments}
The authors would like to acknowledge support from National Science Foundation Grant No. ECCS-1351697 and EFMA-1641099. This work was performed in part at the Cornell NanoScale Science and Technology Facility (CNF), a member of the National Nanotechnology Coordinated Infrastructure (NNCI), which is supported by the National Science Foundation (Grant NNCI-1542081).
\end{acknowledgments}

\appendix
%\counterwithin{figure}{section}
\section{Device Characterization} \label{AppA}
\begin{figure*}
  \includegraphics[scale=0.8]{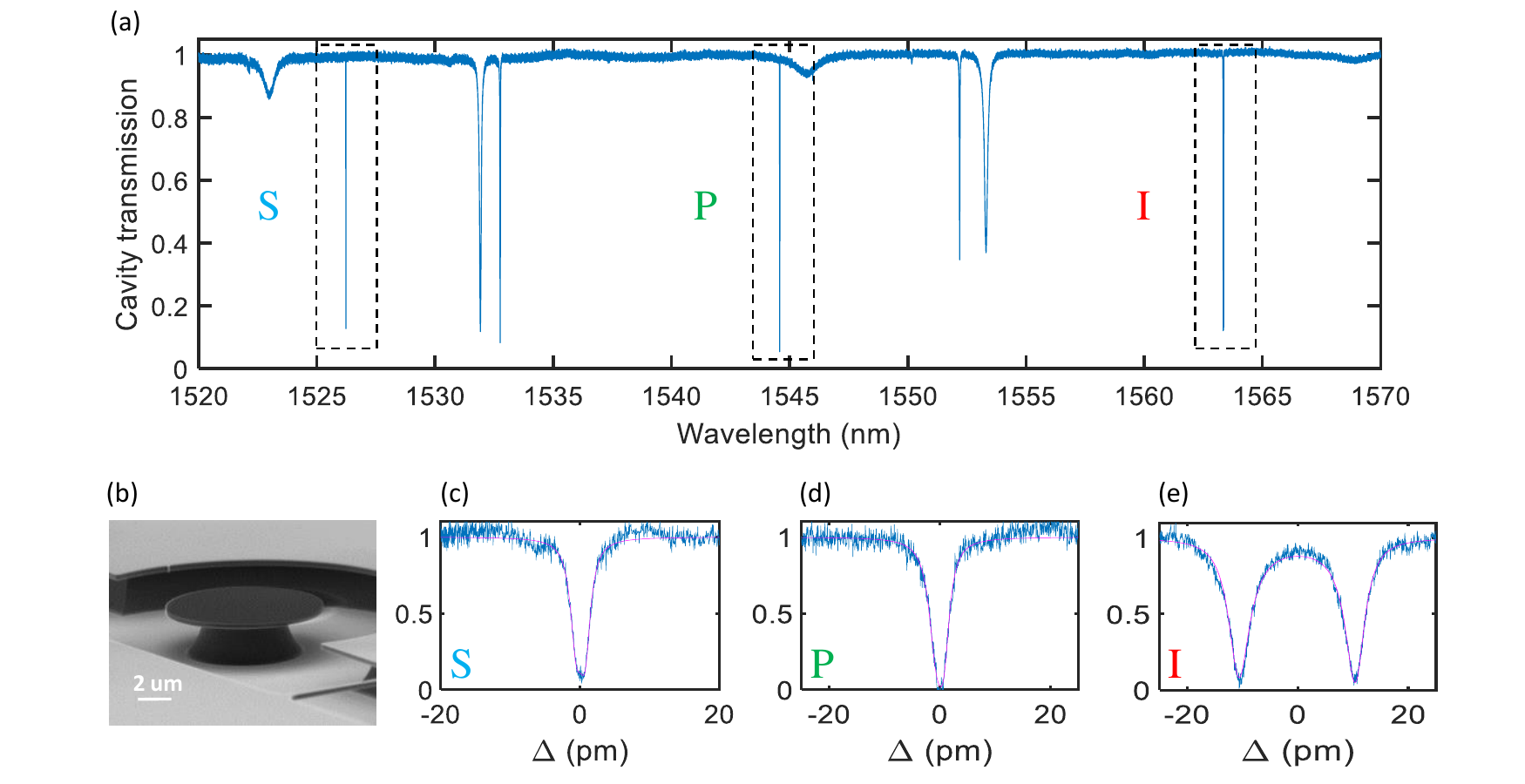}
  \centering
  \caption{(a) Transmission spectrum of resonator A. The signal, pump and idler resonances are identified by dashed boxes and labels S, P and I respectively. (b) SEM image of the device. (c)-(e) Expanded traces of the marked resonances with the fit in magenta. The mean intrinsic quality factors of the two dressed modes and the external quality factors are calculated to be: (c) $1.47 \times 10^6$ and $1.46 \times 10^6$. (d) $1.03 \times 10^6$ and $1.04 \times 10^6$. (e) $1.34 \times 10^6$ and $4.2 \times 10^5$.}
  \label{sfig1}
\end{figure*}

\begin{figure*}
  \includegraphics[scale=1]{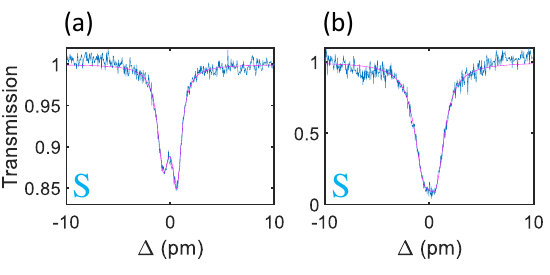}
  \centering
  \caption{(a) Comparison of the signal resonance at different coupling conditions. The external coupling quality factors are: (a) $2 \times 10^7$ and (b) $1.46 \times 10^6$. The 150 MHz (1.16 pm) mode splitting is revealed at a lower external coupling rate in (a)}
  \label{sfig2}
\end{figure*}

Figure \ref{sfig1} shows characterization details for resonator A. A tunable laser is scanned through the device and its transmission is plotted in Fig. \ref{sfig1}(a). Figure \ref{sfig1}(c)-(e) show transmission spectra around the three resonances taking part in the four-wave mixing interaction. An SEM image of the resonator is shown in Fig. \ref{sfig1}(b). The signal (S) resonance has a small mode splitting that becomes visible when weakly coupled to the tapered fiber. Its transmission spectrum at two coupling rates is shown in Fig. \ref{sfig2}.
\par
\begin{figure*}
  \includegraphics[scale=0.8]{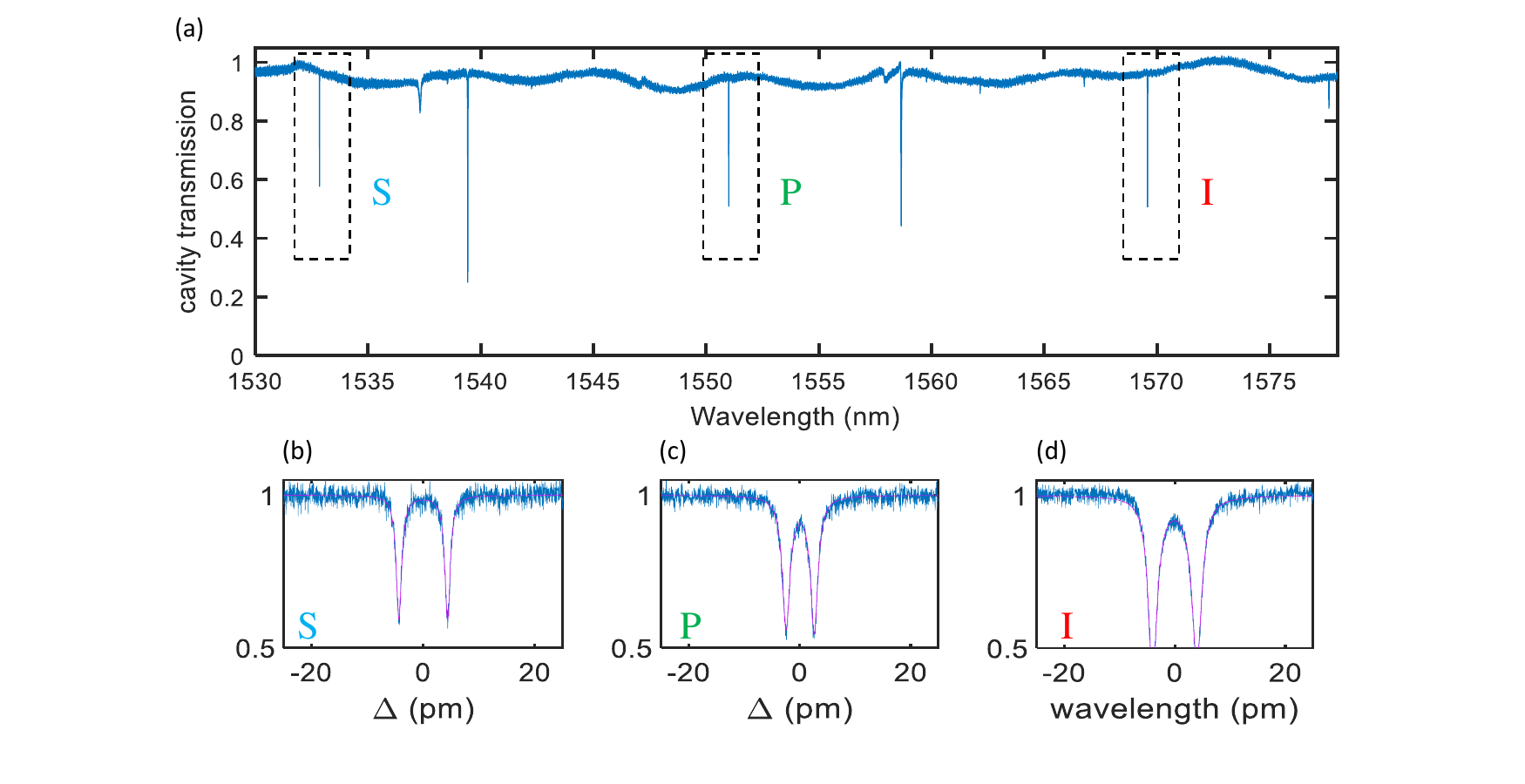}
  \centering
  \caption{(a) Transmission signal of resonator B. The Signal, pump and idler resonances are identified by dashed boxes and labels S, P and I respectively. (b)-(d) Expanded traces of the marked resonances with the fit in magenta. The mean intrinsic quality factors of the two dressed modes and the external quality factors are calculated to be: (b) $1.41 \times 10^6$ and $5.6 \times 10^6$. (c) $1.27 \times 10^6$ and $3.79 \times 10^6$. (d) $1.16 \times 10^6$ and $2.24 \times 10^6$.}
  \label{sfig3}
\end{figure*}

Figure \ref{sfig3} shows the transmission spectrum of resonator B with fitted signal, pump and idler resonances in Fig. \ref{sfig3}(b)-(d). This resonator has mode splitting in all three of its resonances taking part in the interaction.

\bibliography{apssamp}% Produces the bibliography via BibTeX.

\end{document}